\documentclass[11pt]{article}
\usepackage{Blois,epsfig}

\def\be{\begin{equation}}
\def\ee{\end{equation}}
\def\bea{\begin{eqnarray}}
\def\eea{\end{eqnarray}}

\begin{document}
\vspace*{2cm}
\begin{center}
\Large{\textbf{XIth International Conference on\\ Elastic and Diffractive Scattering\\ Ch\^{a}teau de Blois, France, May 15 - 20, 2005}}
\end{center}

\vspace*{2cm}
\title{ANALYTIC PROPERTIES OF HARD EXCLUSIVE AMPLITUDES}

\author{O.V. TERYAEV }

\address{Joint Institute for Nuclear Research,\\
141980 Dubna, Russia}

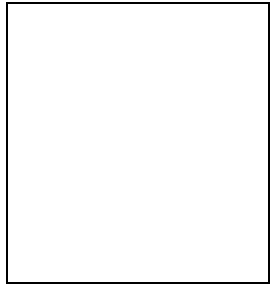
\maketitle\abstracts{
Analytic properties of hard exclusive processes
described by Generalized Parton Distributions (GPD's) are considered.
The analytic continuation of GPD is provided by Generalized Distribution 
Amplitudes (GDA). The GDA's for the production of two $\rho-$mesons 
may give an access to four-quark exotic states.   
The crucial role in the proof of analyticity is played by the Cavalieri
conditions (polynomiality),
resulting in the "holographic" property of GPD, when the full information
about various hard processes
is contained in the one dimensional sections ($x=\pm \xi$)of GPD.
The applicability of analyticity for description of the double diffractive production 
of dileptons and Higgs bosons is discussed. }

\section{Introduction}

Exclusive hard processes described by GPD's are
the subject of extensive theoretical investigations for a few years (see e.g. \cite{BelRad} and Ref. therein).
In the present note we address the problem of analytic properties of the amplitudes 
of Deeply Virtual Compton Scattering (DVCS) and hard meson electroproduction. 
The analytic continuation of GPD involve Generalized Distribution Amplitudes (GDA) 
relevant for the production of meson pairs, in particular, related to exotic four-quark states. 
We analyze the role of polynomiality of GPD's, emerging "holographic" property, and the possible physical 
applications.   

\section{Analyticity in DIS and DVCS}

The leading order contribution of GPD's to the amplitudes of hard processes
-DVCS and (longitudinal) vector meson production is described by the
following integrals:

\be
\label{H}
{\cal H(\xi)}_i=
\int_{-1}^{1}dx H_i(x,\xi)\biggl[
\frac{1}{x-\xi+i\epsilon} \pm \frac{1}{x+\xi-i\epsilon}
\biggr],
\ee
where index $i$ describes the type of GPD, defining also the
choice of $\pm$ sign. The appearance of the same argument $\xi$
in the numerator and denominator results from the zero mass
of the produced photon or its neglecting  for vector meson.
We also
drop the dependence on the momentum transfer $t$.
It is obvious, that only the (anti)symmetric part of
GPD contributes, depending on that sign, which we will only consider
in what follows, so that we will always discuss the
integral,
\be
\label{Hs}
{\cal H(\xi)}=
\int_{-1}^{1}dx \frac {H(x,\xi)}{x-\xi+i\epsilon},
\ee
dropping also the index $i$, as well as (anti)symmetrization index.
This integral looks almost like the dispersion relation with respect to the
variable $s$, where contribution of the crossed channel is usually
taken into account by explicit addition of two terms in (\ref{H})
and the reduction of the integration region to the positive $s$
(corresponding to positive  $x$) only.

There is, however, the notable difference with the forward case,
say, that of Deep Inelastic Scattering. Namely, the numerator depends
also on $\xi$, which prevents from its direct identification as a spectral
density. Nevertheless, the specific properties of $H$ as a Radon transform
makes this dependence inessential.

\section{GPD's in the unphysical region}
Let us first consider the unphysical region $|\xi|>1$.
Note, that the consideration of (\ref{H}) in the unphysical region requires
the appropriate analytical continuation of $H(x,\xi)$.
As it was discussed in detail in \cite{radon}, it is provided by the
integration of double distribution over the straight lines with the
"unphysical" slope:
\be
\label{ParmH}
H(z,\xi)=\int_{-1}^1 dx
\int_{|x|-1}^{1-|x|} dy (F(x,y) + \xi G(x,y)) \delta (z-x-\xi y).
\ee

To identify this object with the physical quantity
one should additionally consider the analytic continuation in $t$.
The resulting object is just GDA \cite{DGPT}. It was recently extended~\cite{GDAV} for the 
case of two vector meson production in the collisions of real and virtual photons. The comparison of charged and neutral $\rho$ production 
identifies~\cite{exotic} the contribution of $I=2$ state, compatible with the existence of exotic 
four-quark resonance.

Note also, that the so-called Polyakov-Weiss (PW) term \cite{PW}, which
originally did not emerge as a Radon transform,
may is also included in such a form (and described by the function $G$),
allowing its consideration in the unphysical region.

\section{Analytic continuation and the role of polynomiality}
 
\subsection{Analyticity in the unphysical region}

Let us consider (\ref{H}) for ($\xi>1$) and
expand the denominator to get:
\begin{eqnarray}
\label{Hs1}
{\cal H(\xi)}=
-\int_{-1}^{1}dx \sum_{n=0}^\infty {H(x,\xi)}\frac {x^n}{\xi^{n+1}}.
\end{eqnarray}
In the forward case, when one have instead of $H$ the forward distribution
which does not depend on $\xi$, this series in the negative
powers of $\xi$ (corresponding to positive powers of $s$) explicitly
manifests  the analyticity of $\cal H$.
In the actual case of GPD's the key role
is played by the mentioned polynomiality property:
the moments of the function $H(x,\xi)$,
namely the integrals in $x$ weighted with $x^n$,
are polynomials of $\xi$ of power $n+1$.
Therefore, the series is still containing only
the non-positive powers of $\xi$,
and the analyticity property is preserved.

\subsection{Dispersion relation, subtractions and holography}

The proven analyticity in the unphysical region allows now to write
the {\it standard} dispersion relation instead of (\ref{Hs}):

\be
\label{Hd}
{\cal H(\xi)}=
\int_{-1}^{1}dx \frac {H(x,x)}{x-\xi+i\epsilon}
\ee

This formula is one of the main results of this paper and
represents the holographic property of GPD: namely,
the full information about, say, DVCS amplitude in the considered leading
approximation is contained in the one-dimensional section $x=\xi$
(related, by the symmetry properties to $x=-\xi$) of the two dimensional
space of $x$ and $\xi$.
In what follows we will study the relations
of dispersion representation with the standard factorization formula:
they either lead to the same result, or, in the case of discrepancy,
the answer coming from the dispersion relations will happen to
be more physically motivated.

Let us consider the difference of (\ref{Hd})
and (\ref{Hs}):

\be
\label{dH}
\Delta {\cal H(\xi)} \equiv
\int_{-1}^{1}dx \frac {H(x,x)-H(x,\xi)}{x-\xi+i\epsilon}
=\sum_{n=1}^{\infty}\frac{1}{n!}
\frac{\partial^n }{\partial \xi^n}\int_{-1}^{1}
H(x,\xi) dx (x-\xi)^{n-1}  =const,
\ee
where we used the Taylor expansion of the first term
$H(x,x)=\sum_{n=0}^{\infty} \frac{1}{n!}\frac{\partial^n }{\partial \xi^n}
H(x,\xi)(x-\xi)^{n}$ and polynomiality
property, interchanging also the order of integration and differentiation.

The emerging constant
term is by no means strange and is nothing else than a subtraction
constant. It is generated by the maximal powers in $\xi$,
provided by PW terms.
To quantify this important relation,
let us calculate $\Delta {\cal H(\xi)}$,
by substituting the definition (\ref{ParmH}): 

\be
\label{sub}
\Delta {\cal H(\xi)}= \int_{-1}^1 dx
\int_{|x|-1}^{1-|x|} dy \frac{G (x,y)}{1-y},
\ee
where we used the following property of delta functions in (\ref{ParmH}): $\delta (z(1-y)-x)= 
\delta (z-x/(1-y))/(1-y) $ (as $|1-y|=1-y$ in the integration region), 
while integrating the $H(z,z)$ term. This proof do not require the existence of infinite 
number of derivatives of $H$ and shows that the very existence of double distribution 
is sufficient to justify the holographic property. so it should be stable against at least LO QCD corrections. 
As we see, only the $G$ function leads to the finite subtraction. 
This provides an extra justification  for the original form of PW term, 
when it resides in the ERBL region $|x| < \xi$. In that case it is obvious, that 
it provides no contribution to the imaginary part of DVCS amplitude,
and is reduced to finite subtraction constant:

\be
\label{Hd3}
\int_{-\xi}^{\xi}dx \frac {D(x/\xi)}{x-\xi+i\epsilon}
=\int_{-1}^{1}dz \frac {D(z)}{z-1}
=const,
\ee

Moreover, the asymptotic form of GPD is also residing at the same region 
but, contrary to PW term, is providing the quadratically growing with energy (like $\xi^{-2}$)
contribution. Such a behaviour is a straightforward counterpart of the two facts:

i) the asymptotic distribution is the non-trivial function only of the ratio $x/\xi$  

ii) the integral $\int dx x H(x.\xi)$ tends to constant due to energy momentum conservation.

As a result, distribution contains the prefactor $\xi^{-2}$, resulting in the energy growth.

\section{Analyticity and double diffraction}
Let us consider the generation of new hard processes by the "substitution" of DA's by GPD's. 
The first stage is just the pion form factor, and the hard meson electroproduction may be considered
as a substitution of one of DA's by GPD's. The  next stage would be the substitution by another GPD 
of the remaining DA, so that one gets the amplitude for Double Diffractive Drell-Yan (DDDY) process $p_1 p_2 \to p^{'}_1 p_2^{'} Q$. 
The explicit calculation of the cross-section in the {\it physical} region, however, results in the 
violation of factorization \footnote{I am indebted to M. Diehl for the illuminating discussion of this problem.}.
Contrary to that, the consideration of the unphysical region $|\xi_{1,2}| >1$, where $\xi_{1,2}=s_{2,1}/s $ and $s_{i}=(p^{'}_i+Q)^2$, 
results in the factorized amplitude 
\be
\label{DDif}
{\cal H(\xi)}=
\int_{-1}^{1}dx dy \frac {H(x,\xi_1)}{x-\xi_1} \frac {H(y,\xi_2)}{y-\xi_2}.
\ee
By the consideration, analogous to the previous section, one may recast it in the form of 
(double and single) subtracted spectral representations. However, the analytic continuation to the physical 
region is now more subtle, as the cuts in $s$ and $s_i$ provide the different signs for the $\i \varepsilon$ addition to $\xi_{1,2}$.    
The symmetric contribution of the combinations of the cuts in $s,s_1$ and $s,s_2$ would lead to the pure real amplitude, 
although the more detailed analysis is required. This method may be also applied to the double diffractive Higgs production,
where heavy quark GPD's of proton should enter. The mechanism of Brodsky, Schmidt and Soffer \cite{soffer} should than correspond 
to the substitution of one of GPD's to the gluon one with corresponding modification of subprocess, being now similar to 
the diffractive meson production at high energies.   

\section*{Conclusions}

The analytic properties of hard exclusive amplitudes are deeply related to the Radon transform properties of GPD's.
The emerging holographic property in the momentum space (complementary to often discussed tomography and holography in 
coordinate space \cite{BelRad})
suggests the new strategy of GPD's studies, focused on its 
extractions at the points $x=\xi$, relevant for imaginary part of amplitude, manifested in Single Spin Asymmetry.
The physical applications of the analyticity and crossing include the search of exotic mesons and double diffractive 
production of dileptons and Higgs bosons.  

\section*{Acknowledgments}
I am grateful to Organizers, and, especially, to Professors Jean Tran Thanh Van and Basarab Nicolescu 
for warm hospitality at Blois and financial support. This work is
partially supported by the grant RFBR 03-02-16816.

\end{document}